\documentclass[preprint,showpacs,showkeys,preprintnumbers,amsmath,amssymb]{revtex4}
 \usepackage{dcolumn}
 \usepackage{bm}

\begin{document}

 \title{A New Route to the Interpretation of Hopf Invariant}

 \author{REN Ji-rong }
 \author{LI Ran }
 \thanks{Electronic mail: liran05@lzu.cn}
 \author{DUAN Yi-shi }
 \affiliation{Institute of Theoretical Physics,  Lanzhou University, Lanzhou, 730000, China}

 \begin{abstract}

 We discuss an object from algebraic topology, Hopf
 invariant, and reinterpret it in terms of the $\phi$-mapping topological
 current theory. The main purpose of this paper is to present a new theoretical
 framework which can directly give the relationship between Hopf
 invariant and the linking numbers of the higher dimensional
 submanifolds of Euclidean space $R^{2n-1}$.
 For the sake of this purpose we introduce a topological tensor current
 which can naturally deduce the $(n-1)$ dimensional topological
 defect in $R^{2n-1}$ space. If these $(n-1)$ dimensional topological
 defects are closed oriented submanifolds of $R^{2n-1}$, they are
 just the $(n-1)$ dimensional knots. The linking number of these
 knots is well defined. Using the inner structure of the
 topological tensor current, the relationship between Hopf
 invariant and the linking numbers of the higher dimensional knots
 can be constructed.
 \end{abstract}



 \maketitle

 \section{Introduction}

 Results from the pure mathematical literature are of little
 difficulty for physicists and can be made accessible to physicists
 by introducing them in common physical methods. In this paper,
 we will discuss an object from algebraic topology, the Hopf
 invariant, and reinterpret it in terms of the $\phi$--mapping topological
 current theory proposed by Duan.
 Hopf studied the third homotopy group of the 2-sphere and showed
 that this group is non-trivial; later Hopf invented more
 non-trivial fibration and finally
 obtained a series of invariants that now bear his name.
 A review article is appeared in Ref.\cite{hopfreview}
 in which the author pointed out the Hopf fibration occurs in at
 least seven different situation in theoretical physics in various
 guises.

 It is well known that Hopf invariant is characterized by the
 homotopy group $\pi_{2n-1}(S^n)$.
 In the case that $n=2$, the
 Hopf invariant $\pi_3(S^2)$ as the linking numbers of knots
 has many applications in condensed matter physics and gauge field theory.
 In fact, we can take the Hopf map $S^3\rightarrow S^2$ as a
 projection in the sense that $S^3$ is a principle fibre bundle
 over the base space $S^2$ with the structure group $U(1)$. The
 standard fibre of Hopf bundle is $S^1$ which is the inverse image of the
 point of $S^2$ under the Hopf map. In 3-sphere, $S^1$ is
 homeomorphous with knot and the topological quantity to describe the
 topological characteristics of these knots(standard fibres) is Hopf invariant.
 As applications, for example,
 a version of nonlinear $O(3)$ $\sigma$-model introduced by
 Faddeev\cite{faddeev}
 allows formation of the stable finite-length solitons which may
 have a form of knot or vortex loop. The topological quantum
 number of these stable defects is just described by the
 Hopf invariant $\pi_3(S^2)$ which is the linking numbers of these knots or vortex loops.
 In gauge field theory, Chern-Simons action have a deep relationship
 with Hopf invariant. In \cite{duanPRD}, by virtue of the decomposition
 of U(1) gauge potential and $\phi$--mapping topological current theory,
 one of authors have
 pointed out that $U(1)$ Chern--Simons action is the
 total sum of all the self-linking and linking numbers of the
 knots inherent in Chern-Simons field theory.

 We \cite{JMPHOPF} have studied the Hopf invariant $\pi_3(S^2)$ by using the
 $\phi-$mapping method. Through the spinor representation of Hopf
 map, the inner topological structure is discovered which indicates that
 Hopf invariant is just the winding number $\pi_3(S^3)$. We also
 obtain a precise expression for Hopf invariant from which one can
 relate the Hopf invariant with the linking number and
 self-linking numbers of Knots.
 In this paper,
 almost using the same method, we will study the general Hopf
 invariant $\pi_{2n-1}(S^n)$ and give a precise expression for it.
 It is revealed that Hopf invariant is the linking numbers of the
 higher dimensional knots or higher dimensional closed oriented
 submanifolds of $R^{2n-1}$. The route to discover the interpretation
 of Hopf invariant appeared in this paper is
 new, at least as far as we know. This will increase our knowledge
 about Hopf invariant without doubt. We have mainly employed the
 tensor and coordinate language which may be nervous for some
 readers. But the result is really perfect. Firstly, we will
 review Hopf invariant according to the general textbook.

 Hopf map\cite{hopfmap} is the
 differentiable map between the spheres of different dimensions
 \begin{equation}
 f:\;S^{2n-1}\rightarrow S^{n}.
 \end{equation}
 Consider a mapping
 \begin{equation}
 n:\;R^{2n-1}\rightarrow S^{n},
 \end{equation}
 which gives a $(n+1)$ dimensional unit vector
 $n^A(A=1,2,\ldots,n+1)$.
 We add the boundary condition
 \begin{equation}
  n(x)|_{x\mapsto \infty}\rightarrow n_0,
 \end{equation}
 where $n_0$ is fixed vector or a fixed point on $S^n$. It is to say that
 we assume infinite distance points of $R^{2n-1}$ is mapped to the same point on $S^n$,
 and therefore the spatial infinity can be
 efficiently contracted to a point, i. e. , $R^{2n-1}\approx S^{2n-1}$. Thus,  the unit
 vector $n(x)$ provides us the mapping $S^{2n-1}\rightarrow S^n$ which is just the Hopf
 map. Under the boundary condition (3),  We will not distinguish $R^{2n-1}$
 with $S^{2n-1}$. So it is intuitively clear to choose Euclidean
 metric in the following discussion.

 In terms of $(n+1)$ dimensional unit vector
 $n^A$, the unit volume element of $S^n$ can
 be constructed like this\cite{moderngeometry}
 \begin{equation}
 \tau=\frac{1}{A(S^n)n!}\epsilon_{A_1A_2\ldots A_{n+1}}n^{A_1}dn^{A_2}\wedge\cdots\wedge
 dn^{A_{n+1}},
 \end{equation}
 which is a closed $n$-form satisfying
 \begin{equation}
 \int_{S^n}\tau=1.
 \end{equation}
 Pulling the unit volume element $\tau$ of $S^n$ back
 to $R^{2n-1}$, $n^*\tau$ must be an exact form. One have
 \begin{equation}
 d\omega=n^*\tau,
 \end{equation}
 where $\omega$ is a $(n-1)$-form on $R^{2n-1}$.
 Hopf invariant\cite{hopfinvariant} which depends only on the homotopy
 group $\pi_{2n-1}(S^n)$ can be written in the form of integral
 \begin{equation}
 H=\int_{S^{2n-1}}\omega\wedge d\omega \;\;.
 \end{equation}
 There are three properties for Hopf invariant\cite{bott}.
 The first, Hopf invariant is independent of the choice
 of $(n-1)$-form $\omega$. This gives us a great convenience in
 the discussion.
 The second, for odd $n$ the Hopf invariant is
 $0$. In this paper, we only consider the case that $n$ is even.
 The third, homotopic maps have the same Hopf invariant. Actually,
 the Hopf invariant is always interpreted as the linking number of
 the preimages $N_k=f^{-1}(p)$ and $N_l=f^{-1}(q)$ of any two
 different values of Hopf mapping $f$.

 Although until now the general Hopf invariant
 have not found as many applications in physics as the Hopf
 invariant $\pi_3(S^2)$ , one must notice in this paper
 that these higher dimensional closed oriented
 submanifolds are deduced naturally from the topological tensor current
 and their generation mechanism is very similar with the topological $p$-branes\cite{pbranes}.
 In string theory, an important aspect is the extra dimension
 theory which implies that spacetime is fundamentally higher dimensional
 with $(4+n)$ spacetime dimensions. Except the usual four dimensional
 Einstein space-time manifold, there exists extra $n$ dimension space which usually shrinks in a
 very small distance. We think that Hopf invariant may be a
 suitable quantity to describe the distorting and linking of the extra dimensional
 space.

 The main purpose in this paper is to present a new theoretical
 framework which directly gives the relationship between Hopf
 invariant and the linking numbers of the higher dimensional knots.
 For this purpose we introduce a topological tensor current
 which can naturally deduce the $(n-1)$ dimensional topological
 defect in $R^{2n-1}$ space. If these $(n-1)$ dimensional topological
 defects are closed oriented submanifolds of $R^{2n-1}$, they are
 just the $(n-1)$ dimensional knots whose linking number
 is well defined. Using the inner structure of the
 topological tensor current, the relationship between Hopf
 invariant and the linking numbers of the higher dimensional knots
 can be constructed.

 \section{inner structure of topological current}

 Recall that the pulling back of $\tau$ is an exact $n$-form and Hopf invariant is independent of the choice of
 $(n-1)$-form $\omega$, so one can construct $(n-1)$-form $\omega$
 like this in terms of the $n$ dimensional unit vector $m^{a}(x)(a=1,2,\ldots,n)$
 \begin{equation}
 \omega=\frac{1}{A(S^{n-1})(n-1)! }
 \epsilon_{a_1a_2\cdots a_n}m^{a_1}dm^{a_2}\wedge\cdots\wedge
 dm^{a_n}.
 \end{equation}
 The unit vector field $m^{a}(x)$ is the function of $R^{2n-1}$,
 where $x$ is the coordinates of $R^{2n-1}$. In fact,
 noticing that $n$ is even, this expression for $\omega$
 which shows the relationship of the Gauss mappings
 between the $n$-dimension and the $(n + 1)$-dimension
 can be directly deduced
 from Eq.(4) and (6) in terms of Gauss-Bonnet-Chern
 theorem\cite{duanzhang} with the help of $(n+1)$-dimensional Euclidean space $R^{n+1}$.

 According to $\phi$-mapping topological current theory\cite{phimap}, $d\omega$
 is of the form of $(n-1)$ order topological current. We introduce a
 topological tensor current
 \begin{eqnarray}
 j^{\mu_1\mu_2\cdots\mu_{n-1}}&=&\frac{1}{(n-1)!}\epsilon^{\mu_1\mu_2\cdots\mu_{2n-1}}
 \partial_{\mu_n}\omega_{\mu_{n+1}\mu_{n+2}\cdots\mu_{2n-1}}\nonumber\\
 &=&\frac{1}{A(S^{n-1})(n-1)!}\epsilon^{\mu_1\mu_2\cdots\mu_{2n-1}}\epsilon_{a_1a_2\cdots a_n}
 \partial_{\mu_n}m^{a_1}\cdots\partial_{\mu_{2n-1}}m^{a_n}.
 \end{eqnarray}
 Obviously the topological tensor current is identically conserved,
 \begin{eqnarray}
 \partial_{\mu_i}j^{\mu_1\mu_2\cdots\mu_{n-1}}=0,\;\;\;i=1,2,\ldots,n-1.
 \end{eqnarray}
 In the $\phi$-mapping theory, the unit vector $m^a(x)$ should be further
 determined by the smooth vectors $\phi^a(x)$, i.e.
 \begin{equation}
 m^a(x)=\frac{\phi^a(x)}{\|\phi\|}\;\;,\;\;\;\;\|\phi\|=\phi^a \phi^a .
 \end{equation}
 In our theory the smooth vector field $\phi^a(x)$ may be looked upon as the generalized
 order parameters for topological defects which is corresponding to the zero points of
 $\phi^a(x)$.

 Substituting Eq.(11) into Eq.(9) and noticing
 \begin{equation}
 dm^a(x)=\frac{1}{\|\phi\|}d\phi^a(x)+\phi^a(x)d\frac{1}{\|\phi\|},
 \end{equation}
 we have
 \begin{equation}
 j^{\mu_1\mu_2\cdots\mu_{n-1}}
 =\frac{1}{A(S^{n-1})(n-1)!}\epsilon^{\mu_1\mu_2\cdots\mu_{2n-1}}\epsilon_{a_1a_2\cdots a_n}
 \partial_{\mu_n}(\frac{\phi^{a_1}}{\|\phi\|^n})
 \partial_{\mu_{n+1}}\phi^{a_2}\cdots\partial_{\mu_{2n-1}}\phi^{a_n}.
 \end{equation}
 Defining the rank-$(n-1)$ Jacobian tensor
 $D^{\mu_1\mu_2\ldots\mu_{n-1}}(\frac{\phi}{x})$ of $\phi$ as
 \begin{equation}
 \epsilon^{\mu_1\mu_2\cdots\mu_{2n-1}}
 \partial_{\mu_{1}}\phi^{a_1}\cdots\partial_{\mu_{n-1}}\phi^{a_n}
 =\epsilon^{a_1a_2\cdots a_n}D^{\mu_1\mu_2\ldots\mu_{n-1}}(\frac{\phi}{x})
 \end{equation}
 and noticing
 \begin{equation}
 \epsilon_{a_1a_2\cdots a_n}\epsilon^{aa_2\cdots
 a_n}=\delta^a_{a_1}(n-1)!,
 \end{equation}
 it follows that
 \begin{equation}
 j^{\mu_1\mu_2\cdots\mu_{n-1}}
 =\frac{1}{A(S^{n-1})}
 \frac{\partial}{\partial\phi^a}(\frac{\phi^{a}}{\|\phi\|^n})
 D^{\mu_1\mu_2\ldots\mu_{n-1}}(\frac{\phi}{x}).
 \end{equation}
 Using the relationship
 \begin{equation}
 {{\phi^a}\over{\left\|\phi\right\|^n }} = \left\{
 \begin{array}{ll}
 - {1 \over {(n- 2)}}{\partial \over {\partial \phi ^a }}\left( {{1
 \over {\left\| \phi \right\|^{n - 2} }}} \right) & \;\;for \;\; n>2  ,\vspace{.4cm} \\
 {\partial \over {\partial \phi ^a }} ln\left\| \phi \right\| &\;\;
 for \;\;n=2,
 \end{array} \right.
 \end{equation}
 and the Green function formula in $\phi$-space
 \begin{equation}
 \Delta _\phi \left( {{1 \over {\left\| \phi \right\|^{n - 2} }}}
 \right) = - \left( {n - 2} \right)A\left( {S^{n - 1} }
 \right)\delta^{(n)}(\phi),
 \end{equation}
 \begin{equation}
 \Delta _\phi \left( ln \left\| \phi \right\| \right) = 2\pi
 \delta^{(2)}(\phi),
 \end{equation}
 where $\Delta _{\phi} =\frac{\partial ^2}{\partial \phi ^a\partial
 \phi ^a} $ is the $d$-dimensional Laplacian operator in $ \phi $
 space, we obtain a $\delta $-function like topological tensor
 current
 \begin{eqnarray}
 j^{\mu_1\mu_2\cdots\mu_{n-1}}=\delta^{(n)}(\phi)D^{\mu_1\mu_2\cdots\mu_{n-1}}(\frac{\phi}{x}),
 \end{eqnarray}
 and find that $j_{\mu_1\mu_2\cdots\mu_{n-1}}\neq 0$ only when $\phi =0$. So, it is essential to
 discuss the solutions of the equations
 \begin{equation}
 \phi ^a(x)=0,\quad a=1,\cdots ,n.
 \end{equation}
 The solution of the above equation plays an crucial role in realization of the
 $\tilde p$--brane scenario\cite{pbranes}.
 Suppose that the vector field $\phi (x)$
 possesses $l$ isolated zeroes, according to the implicit function
 theorem\cite{Goursat}, when the zeroes are regular points
 of $\phi$-mapping, i.e., the rank of the Jacobian matrix
 $[\partial _\mu\phi^a]$ is $n$,
 the solution of equation
 \begin{equation}
 \begin{cases}
 \text{$\phi^1(x^1,x^2,\ldots,x^{2n-1})=0$}, \\
 \text{$\cdots$},\\
 \text{$\phi^n(x^1,x^2,\ldots,x^{2n-1})=0$}.
 \end{cases}
 \end{equation}
 can be  parameterized as
 \begin{eqnarray}
 x^\mu=x^\mu_i(u^1,u^2,\ldots,u^{n-1}),
 \end{eqnarray}
 where $i$ denotes the $i$th solution. Eq.(23) denotes that the
 $(n-1)$-dimensional submanifold $N_k$
 embeds in $R^{2n-1}$ space with $u^I(I=1,2,\cdots,n-1)$
 being the local ordinates of submanifold $N_k$. We assume that $N_k(k=1,2,\cdots,l)$
 are knotted(closed smooth oriented) submanifolds of $R^{2n-1}$.
 It will be seen in the following that Hopf invariant
 is just the linking numbers of these knotted submanifolds.
 For this purpose, we will study the inner structure of
 the topological current, i.e. expand the $n$-dimensional $\delta$-function
 on the $(n-1)$-dimensional singular submanifold $N_k$.

 From the above discussions, we see that the topological current will not disappear only when $\phi =0$.
 Here we will focus on the zero points of order parameter field $\phi$
 and will search for the inner topological structure of
 the topological current. It can be proved that there exists a
 $n$-dimensional submanifold $M_k$ in $R^{2n-1}$ with local parametric
 equation
 \begin{equation}
 x^\mu=x^\mu_i(v^1,v^2,\ldots,v^n),
 \end{equation}
 which is transversal to every $N_i$ at the point $p_i$ with metric
 \begin{equation}\label{normal}
 g_{\mu\nu}B^{\mu}_{I}B^{\nu}_{a}|_{p_i}=0,\quad
 I=1,\cdots ,n-1,\quad a=1,\cdots ,n,
 \end{equation}
 where
 \begin{equation}
 \frac{\partial x^\mu }{\partial u^I}=B^{\mu }_{I}  , \quad
 \frac{\partial x^\nu}{\partial v^a} =B^{\nu }_{a},\quad  \quad  \mu
 ,\nu =1,2,\cdots ,2n-1,
 \end{equation}
 are tangent vectors of $N_i$ and $M_i$ respectively and
 $g_{\mu\nu}=\delta_{\mu\nu}$ for Euclidean space $R^{2n-1}$. As we have
 pointed out in Ref.\cite{DuanMengJMP1993}, the unit vector field
 defined as $m:\partial M_i\rightarrow S^{n-1}$ gives a Gauss map ,
 and the generalized Winding Number can be
 given by the  map
 \begin{equation}\label{}
 W_i  = \frac 1{A(S^{n-1})(n-1)!}\int_{\partial M_i}m^{*}(\varepsilon
 _{A_1\cdots A_n}m^{A_1}dm^{A_2}\wedge \cdots \wedge dm^{A_n}),
 \end{equation}
 where $m^*$ denotes the pull back of map $m$ and  $\partial M_i$ is
 the boundary of the neighborhood $M_i$ of $p_i$ on $R^{2n-1}$ with
 $p_i\notin \partial M_i,$ $M_i\cap M_j=\emptyset $. It means that,
 when the point $v^b$ covers $\partial M_i$ once,
 the unit vector $m$ will cover the unit sphere
 $S^{n-1}$ for $W_i$ times. Using the Stokes' theorem in exterior
 differential form and  duplicating the derivation of
 ((9) from (20)), we obtain
 \begin{equation} \label{wind}
 W_i=\int_{M_i}\delta (\phi (v))D(\frac \phi v)d^nv,
 \end{equation}
 where $D(\frac \phi v)$ is the usual Jacobian determinant of $\vec
 \phi $ with respect to $v$
 \begin{equation}
 \varepsilon ^{A_1\cdots A_n}D(\frac \phi v)=\varepsilon ^{\mu
 _1\cdots \mu _n}\partial _{\mu _1}m^{A_1}\partial _{\mu
 _2}m^{A_2}\cdots \partial _{\mu _n}m^{A_n}.
 \end{equation}
 According to the $\delta $-function theory\cite{Schouten},
 we know that $\delta (\phi)$ can be
 expanded as
 \begin{equation}
 \delta (\phi )=\sum_{i=1}^l c_i \delta (N_i),
 \end{equation}
 where the coefficients $c_i$ must be positive, i.e., $c_i=|c_i|$.
 $\delta (N_i)$ is the $\delta $--function on a submanifold
 $N_i$ in $R^{2n-1}$ which had been given in Ref\cite{DuanMengJMP1993}
 \begin{equation}
 \delta (N_i)=\int_{N_i}\delta ^{(2n-1)}(x-x_i(u))\sqrt{g_u}d^{(n-1)}u,
 \end{equation}
 where $g_u=det(g_{IJ})$ is the metric of submanifold $N_i$.
 Substituting Eq.(31) and Eq.(32) into Eq.(28),
 and calculating the integral, we get the expression of $c_i$,
 \begin{equation}\label{}
 c_i=\frac{\beta _i}{\left|J \left(\frac{\phi
 }{v}\right)\right|_{p_i}}=\frac{\beta _i\eta _i}{\left. J
 \left(\frac{\phi }{v}\right)\right|_{p_i}},
 \end{equation}
 where the positive integer $\beta _i=|W_i|$ is called the Hopf index
 of $\phi $--mapping on $M_i$, and $\eta _i=sgn(J(\frac \phi
 v))|_{p_i}=\pm 1$ is the Brouwer degree\cite{DuanMengJMP1993}.
 So we find the relations between the Hopf index $\beta _i,$ the
 Brouwer degree $ \eta _i$, and the winding number $W_i$
 \begin{equation}
 W_i=\beta _i\eta _i.
 \end{equation}

 From Eq.(20), we have
 \begin{eqnarray}
 j^{\mu_1\mu_2\cdots\mu_{n-1}}
 &=&\delta^{(n)}(\phi)D^{\mu_1\mu_2\cdots\mu_{n-1}}(\frac{\phi}{x})\nonumber\\
 &=&\sum_{k=1}^{l}W_k\int_{N_k}\frac{D^{\mu_1\mu_2\cdots\mu_{n-1}}(\frac{\phi
 }{x})}{D(\frac{\phi}{v})}|_{N_i}
 \delta^{(2n-1)}(x-x_k(u))\sqrt{g_u}d^{(n-1)}u
 \end{eqnarray}
 Since on the singular submanifold $N_i$ we have
 \begin{equation}
 \phi^a(x)|_{N_i}=\phi^a(x_i^1(u),\cdots,x_i^{(2n-1)}(u))|_{N_i}=0,
 \end{equation}
 which lead to
 \begin{equation}
 \partial_\mu\phi^a\frac{\partial x^\mu}{\partial u^I}=0.
 \end{equation}
 Using this expression, one can prove the relation between two
 Jacobian
 \begin{equation}
 D^{\mu_1\mu_2\cdots\mu_{n-1}}(\frac{\phi }{x})|_{N_i}=
 \epsilon_{I_1I_2\cdots I_{n-1}}
 \frac{\partial x^{\mu_1}}{\partial u^{I_1}}\cdots\frac{\partial x^{\mu_{n-1}}}{\partial u^{I_{n-1}}}
 D(\frac{\phi}{v}).
 \end{equation}
 Therefore, the inner topological structure of the topological current
 can be ultimately expressed as
 \begin{eqnarray}
 j^{\mu_1\mu_2\cdots\mu_{n-1}}=\sum_{k}W_k\int_{N_k}\epsilon_{I_1I_2\cdots
 I_{n-1}}\frac{\partial x^{\mu_1}}{\partial u^{I_1}}\cdots\frac{\partial x^{\mu_{n-1}}}{\partial u^{I_{n-1}}}
 \delta^{(2n-1)}(x-x_k(u))\sqrt{g_u}d^{(n-1)}u.
 \end{eqnarray}
 From this equation, we conclude that the inner structure of
 the topological current $j^{\mu_1\mu_2\cdots\mu_{n-1}}$ is labelled by the
 total expansion of $\delta(\phi)$, which characters the $l$
 submanifolds $N_k$ with the quantized topological charge $W_k$.

 \section{Hopf invariant}

 Now, we will discuss Hopf invariant in terms of
 the inner structure of the topological current
 obtained in the last section.

 In terms of Eq.(9) Hopf invariant can be written as
 \begin{eqnarray}
 H=\frac{1}{(n-1)!}\int\omega_{\mu_1\mu_2\cdots\mu_{n-1}}
 j^{\mu_1\mu_2\cdots\mu_{n-1}}d^{(2n-1)}x.
 \end{eqnarray}
 Substituting Eq.(37) into above equation, Hopf invariant is in the form
 \begin{equation}
 H=\frac{1}{(n-1)!}\sum_{k}W_k\int_{N_k}\epsilon_{I_1I_2\cdots
 I_{n-1}}\frac{\partial x^{\mu_1}}{\partial u^{I_1}}\cdots\frac{\partial x^{\mu_{n-1}}}{\partial u^{I_{n-1}}}
 \omega_{\mu_1\mu_2\cdots\mu_{n-1}}\sqrt{g_u}d^{(n-1)}u,
 \end{equation}
 or the following form
 \begin{equation}
 H=\sum_{k}W_k\int_{N_k}\omega,
 \end{equation}
 where $N_k$ is a knotted(closed smooth oriented) submanifold of $R^{2n-1}$.  Hopf invariant does
 not depend on the selection of the $(n-1)$-form $\omega$.
 It is difficult to continue the discussion directly from Eq.(40) or Eq.(41).
 But we find that it is useful to express $\omega$ in terms of the
 topological current. For this purpose ,we introduce a new tensor
 \begin{equation}
 C_{\mu_1\mu_2\cdots\mu_{n-1}}=
 \epsilon_{\mu_1\mu_2\cdots\mu_{2n-1}}
 \partial_{\mu_n}j^{\mu_{n+1}\mu_{n+2}\cdots\mu_{2n-1}}.
 \end{equation}
 If we impose the condition which is very similar with the selection of
 Columb gauge in classical electrodynamics
 \begin{equation}
 \partial_{\mu_i}\omega_{\mu_1\mu_2\cdots\mu_{n-1}}=0,\;\;\;i=1,2,\ldots,n-1.
 \end{equation}
 one can easily get
 \begin{equation}
 C_{\mu_1\mu_2\cdots\mu_{n-1}}=(n-1)!\partial_{\mu}\partial_{\mu}\omega_{\mu_1\mu_2\cdots\mu_{n-1}}.
 \end{equation}
 Using the Green function formula
 \begin{equation}
 \Delta_{(2n-1)} \left( {{1 \over {\left\| x-y \right\|^{2n - 3} }}}
 \right) = - \left( {2n - 3} \right)A\left( {S^{2n - 2} }
 \right)\delta^{(2n-1)}(x-y),
 \end{equation}
 where $\Delta_{(2n-1)}=\partial_\mu\partial_\mu$ is
 the $(2n-1)$-dimensional Laplacian operator in $R^{2n-1}$,
 the general formal solution of Eq.(44) is
 \begin{eqnarray}
 \omega_{\mu_1\mu_2\cdots\mu_{n-1}}(x)&=&
 \frac{1}{(2n-3)A(S^{2n-2})(n-1)!}\int d^{(2n-1)}y
 \frac{C_{\mu_1\mu_2\cdots\mu_{n-1}}(y)}{\|x-y\|^{2n-3}}\nonumber\\
 &=&\frac{1}{(2n-3)A(S^{2n-2})(n-1)!}
 \epsilon_{\mu_1\mu_2\cdots\mu_{2n-1}}\int d^{(2n-1)}y
 \frac{\partial_{\mu_n}j^{\mu_{n+1}\mu_{n+2}\cdots\mu_{2n-1}}(y)}{\|x-y\|^{2n-3}}.
 \end{eqnarray}
 Because under the boundary condition (3) the topological
 current disappears on the boundary, one can easily get
 \begin{eqnarray}
 \omega_{\mu_1\mu_2\cdots\mu_{n-1}}(x)&=&\frac{1}{(2n-3)A(S^{2n-2})(n-1)!}\nonumber\\
 &&\times\epsilon_{\mu_1\mu_2\cdots\mu_{2n-1}}\int d^{(2n-1)}y
 j^{\mu_{n+1}\mu_{n+2}\cdots\mu_{2n-1}}(y)\partial_{\mu_n}\frac{1}{\|x-y\|^{2n-3}}.
 \end{eqnarray}
 Substituting it into Eq.(39), one get
 \begin{eqnarray}
 H&=&\frac{1}{(2n-3)A(S^{2n-2})[(n-1)!]^2}
 \epsilon_{\mu_1\mu_2\cdots\mu_{2n-1}}\int d^{(2n-1)}x
 j^{\mu_1\mu_2\cdots\mu_{n-1}}(x)\nonumber\\
 &&\times\int d^{(2n-1)}y j^{\mu_{n+1}\mu_{n+2}\cdots\mu_{2n-1}}(y)
 \partial_{\mu_n}\frac{1}{\|x-y\|^{2n-3}}.
 \end{eqnarray}
 According to the inner structure of topological current (38), one get
 \begin{eqnarray}
 H&=&\frac{1}{(2n-3)A(S^{2n-2})[(n-1)!]^2}
 \sum_{k,l}W_kW_l\epsilon_{\mu_1\mu_2\cdots\mu_{2n-1}}
 \int_{N_k}\epsilon_{I_1I_2\cdots I_{n-1}}
 \frac{\partial x^{\mu_1}}{\partial u^{I_1}}\cdots
 \frac{\partial x^{\mu_{n-1}}}{\partial u^{I_{n-1}}}\sqrt{g_u}d^{(n-1)}u\nonumber\\
 &&\times\int_{N_l}\epsilon_{J_1J_2\cdots J_{n-1}}\frac{\partial y^{\mu_{n+1}}}{\partial v^{J_1}}\cdots
 \frac{\partial y^{\mu_{2n-1}}}{\partial v^{J_{n-1}}}\sqrt{g_v}d^{(n-1)}v
 \partial_{\mu_n}\frac{1}{\|x-y\|^{2n-3}}.
 \end{eqnarray}
 Note that the integral is on the submanifolds $N_k$ of $R^{2n-1}$,  this
 gives the connection between Hopf invariant and the
 submanifolds $N_k$. To reveal the relationship
 between Hopf invariant and linking number of higher dimensional knots,
 we must seek a formula for the linking
 number of submanifolds of Euclidean space $R^{2n-1}$.

 \section{linking number of higher dimensional knot}

 Recall the definition of higher dimensional knot, in knot theory, a
 n-knot $K$ is a knotted(closed, oriented) n-manifold. The strict definition of
 n-knot need the knowledge of Seifert surface. We are not intending to
 discuss the strict definition of higher dimensional knot. We are
 interested in two closed smooth oriented manifolds of the same
 dimensions $(n-1)$ which is embedded in $R^{2n-1}$ space\cite{definelink}. Then the
 intersection number or linking number of the two manifolds is well
 defined. In the classical case, knot,i.e. one dimensional knot $\gamma$,
 is in fact an embedding map to $R^3$
 \begin{eqnarray}
 \gamma:\;\;\;S^1\rightarrow R^3.
 \end{eqnarray}
 The well known characteristic number for describes the topology
 of  knots is linking number which is defined as the degree of map
 $S^1\times S^1\rightarrow S^2$. It is easy to generalize the ideas in
 the classical case into higher dimensional knot.

 From Whitney's embedding theorem\cite{moderngeometry}, we know that any smooth
 connected closed manifold $M$ of dimension $n$ can be smoothly
 embedded in $R^{2n+1}$.Consider two closed smooth
 oriented manifold $N_k$ and $N_l$ of
 the same dimensions $(n-1)$, the continuous mappings embedding $N_k$ and $N_l$
 into $R^{2n-1}$ that $N_k$ and $N_l$ do not intersect is
 \begin{eqnarray}
 &N_k&:\;\;x^\mu=x^\mu(u^1,u^2.\ldots,u^{n-1}),\nonumber\\
 &N_l&:\;\;y^\mu=y^\mu(v^1,v^2,\ldots,v^{n-1}),
 \end{eqnarray}
 where $u^I,v^J(I,J=1,2,\ldots,n-1)$ are the local parameters of
 $N_k$ and $N_l$ respectively.
 Consider the cartesian $N_k\times N_l$ given the canonical
 orientation. We define a map
 \begin{eqnarray}
 e:\;\;\;N_k\times N_l\rightarrow S^{2n-2},
 \end{eqnarray}
 by associating to each $(u,v)\in N_k\times N_l$ the unit vector is
 \begin{eqnarray}
 e^\mu(u,v)=\frac{x^\mu(u)-y^\mu(v)}{\|x-y\|},\;\;\mu=1,2,\cdots,2n-1.
 \end{eqnarray}
 The degree of this map is the linking number $Lk(N_k,N_l)$. The
 volume element of the $(2n-2)$ sphere $S^{2n-2}$ is
 \begin{eqnarray}
 d\sigma=\frac{1}{A(S^{2n-2})(2n-2)!}\epsilon_{\mu_1\mu_2\ldots\mu_{2n-1}}e^{\mu_1}de^{\mu_2}\wedge\cdots\wedge
 de^{\mu_{2n-1}}.
 \end{eqnarray}
 Then, clearly
 \begin{eqnarray}
 {\cal{L}}k(N_k,N_l)=deg\;e=\int_{N_k}\int_{N_l}e^*d\sigma.
 \end{eqnarray}
 The linking number ${\cal{L}}k(N_k,N_l)$ is integer-valued and remains
 constant under the continuous varying of embedding mapping.
 Noticing that
 \begin{eqnarray}
 de^\mu=\frac{dx^\mu-dy^\mu}{\|x-y\|}+(x^\mu-y^\mu)d\frac{1}{\|x-y\|},
 \end{eqnarray}
 the volume element is
 \begin{eqnarray}
 d\sigma&=&\frac{1}{A(S^{2n-2})(2n-2)!}\epsilon_{\mu_1\mu_2\ldots\mu_{2n-1}}\frac{(x^{\mu_1}-y^{\mu_1})}{\|x-y\|^{2n-1}}
 (dx^{\mu_2}-dy^{\mu_2})\wedge\cdots\wedge(dx^{\mu_{2n-1}}-dy^{\mu_{2n-1}})\nonumber\\
 &=&\frac{1}{A(S^{2n-2})(2n-2)!}\epsilon_{\mu_1\mu_2\ldots\mu_{2n-1}}\frac{(x^{\mu_1}-y^{\mu_1})}{\|x-y\|^{2n-1}}
 (\frac{\partial x^{\mu_2}}{\partial u^{I^2}}du^{I^2}-\frac{\partial y^{\mu_2}}{\partial v^{J^2}}dv^{J^2})
 \wedge\cdots\nonumber\\&&\wedge(\frac{\partial x^{\mu_{2n-1}}}{\partial u^{I^{2n-1}}}du^{I^{2n-1}}
 -\frac{\partial y^{\mu_{2n-1}}}{\partial v^{J^{2n-1}}}dv^{J^{2n-1}}).
 \end{eqnarray}
 The linking number $Lk(N_k,N_l)$ is ultimately expressed as
 \begin{eqnarray}
 {\cal{L}}k(N_k,N_l)&=&\frac{1}{(2n-3)A(S^{2n-2})[(n-1)!]^2}\epsilon_{\mu_1\mu_2\ldots\mu_{2n-1}}
 \int_{N_k}\epsilon_{I_1I_2\cdots I_{n-1}}
 \frac{\partial x^{\mu_1}}{\partial u^{I_1}}\cdots
 \frac{\partial x^{\mu_{n-1}}}{\partial u^{I_{n-1}}}\sqrt{g_u}d^{(n-1)}u\nonumber\\
 &&\times\int_{N_l}\epsilon_{J_1J_2\cdots J_{n-1}}\frac{\partial y^{\mu_{n+1}}}{\partial v^{J_1}}\cdots
 \frac{\partial y^{\mu_{2n-1}}}{\partial v^{J_{n-1}}}\sqrt{g_v}d^{(n-1)}v
 \partial_{\mu_n}\frac{1}{\|x-y\|^{2n-3}}.
 \end{eqnarray}
 At last, comparing the above expression with Eq.(48) one get Hopf invariant
 \begin{eqnarray}
 H=\sum_{k,l}W_kW_l {\cal{L}}k(N_k,N_l).
 \end{eqnarray}
 This expression for Hopf invariant is consistent with the
 interpretation of Hopf invariant from mathematical literature\cite{bott}
 which is the linking number of
 the preimages $N_k=f^{-1}(p)$ and $N_l=f^{-1}(q)$ of any two
 different values of Hopf mapping $f$.

 At last, to complete this paper, a final remark is necessary.
 Following Alberto S. Cattaneo's work on $\textbf{BF}$
 topological field theory\cite{bftheory},
 similarly, we can construct a generalized abelian field theory of Hopf invariant's
 type which is defined by the action functional
 \begin{equation}
 S(A)=\int_M A\wedge dA,
 \end{equation}
 where $A$ is an ordinary $(n-1)$-form on an $(2n-1)$-dimensional
 manifold $M$ with $n$ being even.
 The basic field or independent field variable
 is the $(n-1)$ order tensor field $A_{\mu_1\cdots\mu_{n-1}}$
 which probably describes the high spin particles.
 The generalization of abelian gauge symmetries is in
 this case given by transformations of the form $A\mapsto A+d\sigma,
 \sigma\in\Omega^{n-2}(M)$. As pointed out in this paper,
 the explanation of the action (60) is an
 topological invariant which in the case $M=R^{2n-1}$ turns
 out to be a function of the linking number of the
 higher-dimensional knot. The classical Euler--Lagrange
 equations of motion for the action
 gives an identical equation. As in Chern-Simons theory,
 field variable $A_{\mu_1\cdots\mu_{n-1}}$ should be coupled with the matter field
 or the matter current to display
 the property of the action.
 In quantum field theory, in addition to the action functional, one also wishes to
 pick a suitable class of gauge invariant observables. As done by
 E.Witten in his pioneer work\cite{witten} which gives a theoretical
 framework to study the
 relationship between Chern-Simons action and the knot invariant,
 the obvious generalization of a Wilson loop as a observables has the form
 \begin{equation}
 W(N_i)=exp\{\frac{i}{\hbar}\lambda\int_{N_i} A\},
 \end{equation}
 where $N_i$ is a knotted(closed oriented) $(n-1)$-dimensional submanifolds of $M$
 and $\lambda$ is the coupling constant. Then we propose the Feynman
 path integral
 \begin{equation}
 \int{\cal{D}}A\;exp\{\frac{i}{\hbar}S\}\prod_{i=1}^{l}W(N_i).
 \end{equation}
 The generalization of Wilson loops
 where the connection is replaced by a form $A$ of higher
 degree and the loop by a higher-dimensional submanifold is then
 natural and might have applications to the theories of $\tilde
 p$--brane. In terms of the $\phi$--mapping
 topological current theory, $(n-1)$ dimensional submanifold can be
 naturally deduced from the action (60), which plays a crucial role
 in realization of the $\tilde p$--brane scenario. If the action (60)
 is significative, a further study of the generalized abelian
 field theory and the application to
 $\tilde p$--brane will attract our future attention.

 \section*{ACKNOWLEDGEMENT}

 This work was supported by the National Natural Science Foundation
 of China.

 \end{document}